# Bottom-Up Perspectives on AI Governance: Insights from User Reviews of AI Products

Stefan Pasch[1]

## Abstract

With the growing importance of AI governance, numerous high-level frameworks and principles have been articulated by policymakers, institutions, and expert communities to guide the development and application of AI. While such frameworks offer valuable normative orientation, they may not fully capture the practical concerns of those who interact with AI systems in organizational and operational contexts. To address this gap, this study adopts a bottom-up approach to explore how governance-relevant themes are expressed in user discourse. Drawing on over 100,000 user reviews of AI products from G2.com, we apply BERTopic to extract latent themes and identify those most semantically related to AI governance. The analysis reveals a diverse set of governance-relevant topics spanning both technical and non-technical domains. These include concerns across organizational processes—such as planning, coordination, and communication—as well as stages of the AI value chain, including deployment infrastructure, data handling, and analytics. The findings show considerable overlap with institutional AI governance and ethics frameworks on issues like privacy and transparency, but also surface overlooked areas such as project management, strategy development, and customer interaction. This highlights the need for more empirically grounded, user-centered approaches to AI governance—approaches that complement normative models by capturing how governance unfolds in applied settings. By foregrounding how governance is enacted in practice, this study contributes to more inclusive and operationally grounded approaches to AI governance and digital policy.

## 1. Introduction

As artificial intelligence (AI) systems become increasingly embedded in organizational life and everyday technologies, questions of how these systems are governed, regulated, and held accountable have gained critical importance. From algorithmic decision-making in hiring and finance to generative AI tools used by knowledge workers, AI now affects a broad range of stakeholders — including developers, policymakers, business leaders, and end-users. As a result, AI governance has emerged as a central concern across industry, academia, and public policy, reflecting growing awareness of the risks, responsibilities, and policy challenges associated with AI use (Yeung, 2018; Brundage et al., 2020). In response, numerous frameworks have been proposed to guide the ethical, responsible, and trustworthy development and deployment of AI technologies. These include influential documents such as the European Commission's *Ethics Guidelines for Trustworthy AI* (2019), the *OECD AI Principles* (2019), and a growing array of national strategies and corporate codes (Fjeld et al., 2020; Jobin et al., 2019). Across these frameworks, a shared set of principles has emerged — emphasizing privacy, transparency, fairness, accountability, human oversight, and societal benefit.

[1] Division of Social Science & AI, Hankuk University of Foreign Studies: stefan.pasch@outlook.com

Yet despite the prominence of these frameworks, most efforts to define and structure AI governance and ethics remain top-down in nature. They are typically created by policymakers, institutional experts, or synthesized through academic literature reviews, rather than through engagement with those who use or manage AI in applied settings. As Mittelstadt (2019) and Sigfrids et al. (2023) have argued, there is a pressing need to complement these expert-driven perspectives with bottom-up insights. Governance unfolds not only in legal norms and high-level principles but also in day-to-day sociotechnical practices and user interactions (Leikas et al., 2019).

Employees integrating AI into workflows, managers overseeing automation strategies, and customers responding to AI-driven interfaces may emphasize governance concerns that differ from established ethical principles. These can include usability, access management, internal communication, or strategic alignment. dimensions of governance that often go unrecognized in formal guidelines (Wagner, 2018; Leikas et al., 2019). This creates a growing gap between what AI governance frameworks prescribe and what users actually encounter "on the ground".

Understanding these bottom-up concerns is not only a matter of analytical completeness, but also of practical policy relevance. If governance frameworks overlook the lived experiences of those interacting with AI systems in applied contexts, they risk being misaligned with real-world needs—limiting their legitimacy, uptake, and effectiveness. For policymakers, surfacing these user-level signals is essential for designing adaptive, responsive, and context-aware governance mechanisms that extend beyond abstract principles into operational reality (Vaele & Bass 2019; Larsson, 2021).

In response to this gap, this study adopts a bottom-up, empirical approach to explore how users themselves articulate and experience AI governance. Rather than beginning with predefined categories, we use BERTopic—a transformer-based topic modeling method—to analyze over 100,000 user reviews of AI tools from the G2.com software platform. We then identify governance-relevant topics by using semantic similarity techniques to extract themes that reflect how everyday users articulate concerns related to AI governance.

The findings reveal a rich landscape of concerns, many of which go beyond the boundaries of existing governance frameworks. Alongside expected topics such as privacy and accountability, users frequently raise issues related to project management, strategic planning, organizational structure, customer support, and product communication. These insights suggest that AI governance is best understood not solely as a matter of ethics or compliance, but as a dynamic, sociotechnical process distributed across technical systems, organizational structures, and end-user interactions (Amoore, 2020; Sigfrids et al., 2023).

By doing so, this paper contributes to the AI governance literature in three important ways. First, it introduces a replicable methodological framework for identifying governance-relevant topics using semantic similarity and unsupervised topic modeling, which can support future empirical work in the field. Second, it offers a novel, bottom-up mapping of governance-related concerns based on user-generated discourse. This sheds light on how governance is perceived and experienced in practical, organizational settings—beyond the abstract language of principles. Third, by comparing bottom-up concerns with official frameworks, it identifies both alignments and blind spots—especially in areas such as deployment infrastructure, support ecosystems, and strategic coherence. In doing so, the paper argues for a more integrated and participatory model of AI governance, one that draws on user experience to enhance the legitimacy, responsiveness, and practical efficacy of digital policy frameworks.

## 2. Literature Review

### 2.1. AI Governance

The term *AI governance* has become increasingly central to discussions about the social, ethical, and institutional implications of artificial intelligence. Yet despite its widespread use, the concept lacks a clear and consistent definition—creating challenges not only for scholarly precision but also for policy implementation. In the absence of definitional agreement, efforts to regulate or benchmark AI systems often depend on ambiguous or overlapping normative ideals, which can obscure accountability and hinder operationalization in real-world settings. Scholars and policymakers often use related terms such as *ethical AI*, *trustworthy AI*, or *responsible AI*, reflecting overlapping but distinct concerns (Jobin et al., 2019; Mittelstadt, 2019). This conceptual ambiguity has led to calls for clearer delineation of what governance entails in the context of AI technologies.

One influential definition comes from Mäntymäki et al. (2022, p.604), who describe AI governance as "a system of rules, practices, processes, and technological tools that are employed to ensure an organization's use of AI technologies aligns with the organization's strategies, objectives, and values; fulfills legal requirements; and meets principles of ethical AI followed by the organization." This framing situates governance at the intersection of normative ideals and organizational implementation, emphasizing both compliance and strategic alignment. In line with this, other scholars have emphasized the sociotechnical nature of AI governance, noting that it emerges not only through formal regulation but also through day-to-day infrastructural and operational practices (Bogen & Winecoff, 2024; Raji et al., 2020). Governance, in this view, is shaped as much by system design decisions, deployment routines, and team workflows as by external policies or ethical codes.

While this broader framing is gaining traction, most existing frameworks for AI governance remain grounded in top-down, principle-based models. These include the European Commission's *Ethics Guidelines for Trustworthy AI* (2019), the *OECD AI Principles* (2019), and a range of national strategies and corporate codes. Although often labeled as "ethical" or "responsible" AI guidelines, these documents serve as de facto governance instruments: they inform institutional standards, guide regulatory efforts, and structure assessments of AI system design and deployment (Larsson, 2021; Reinhardt, 2023; Papagiannidis et al., 2025). As Larsson (2021) argues, the prominence of such ethical guidelines reflects both their practical utility and the regulatory vacuum they fill — emerging in a context where legal and enforceable governance mechanisms remain underdeveloped. This dominance of soft law instruments hints at the absence of comparable, binding AI governance frameworks, especially in fast-evolving domains where regulatory lag is significant. Their influence nonetheless extends across sectors, shaping how AI governance is defined and enacted in real-world contexts. Across these frameworks, a recurring set of normative themes can be observed — including privacy, transparency, accountability, fairness, human oversight, and societal benefit (Fjeld et al., 2020; Jobin et al., 2019). These principles reflect broad expectations about how AI should operate in society and frequently serve as anchor points for discussions of trustworthy or legitimate AI governance. As such, even when not explicitly labeled as governance, these frameworks have become central to how governance is conceptualized, benchmarked, and implemented.

## 2.2. The Need for Bottom-Up Perspectives on AI Governance

While frameworks such as the EU's *Ethics Guidelines for Trustworthy AI* have received wide attention in policy, industry, and academic debates (Larsson, 2021), they have also come under increasing scrutiny for their limited applicability in real-world contexts. These documents are typically developed by institutional experts, policymakers, and regulators — rather than by those who use, manage, or are directly affected by AI systems in applied settings. As a result, they often reflect high-level ideals rather than the day-to-day governance concerns encountered by practitioners or end-users. Scholars have noted that such principles often remain too abstract, vague, or removed from the realities of how AI systems are built, deployed, and experienced (Mittelstadt, 2019; Schiff et al., 2021; Morley et al., 2021). Moreover, the key principles identified by policymakers and experts may not fully capture the dimensions that everyday users of AI systems care about or actively engage with. This disconnection risks weakening the effectiveness, legitimacy, and adoption of AI governance frameworks, particularly when they are translated into policy or institutional standards.

A growing body of scholarship emphasizes that AI governance is not only a matter of formal regulation but also of interaction with AI systems in daily use. Bogen & Winecoff (2024) and Raji et al. (2020) describe governance as a sociotechnical process shaped by infrastructure, organizational routines, and user practices. From this perspective, governance is not simply imposed through rules or codes but emerges through the ways people implement, interpret, and work around AI in practice.

In fact, users and practitioners often act as de facto governance agents. Through feedback mechanisms, usage patterns, internal workarounds, or even resistance to certain systems, they help shape how AI functions within their organizations (Stark & Hoffmann, 2019). These interactions represent a lived form of governance — one that unfolds across roles, departments, and contexts rather than through formal regulation alone. This shift expands the scope of governance beyond the developer–regulator dyad, calling attention to a broader ecosystem of stakeholders. As Sigfrids et al. (2023) emphasize, AI affects a wide range of actors — including employees, managers, customers, and those at the margins of development. Making these voices visible is not only a matter of inclusivity but also essential for designing holistic governance frameworks.

Bringing these user-centered perspectives into AI governance is especially important not only because top-down models may face implementation challenges — but because they risk overlooking entire categories of concern. While abstract values like fairness or transparency dominate ethical discourse, they may not fully reflect the concerns of those interacting with AI systems in applied settings. Users embedded in organizational and operational contexts often face challenges that are more immediate and practice-oriented. These may include the integration of AI tools into existing workflows, the coordination of human–machine tasks, or the implications of automation for internal roles and responsibilities. These themes — often central to how AI is experienced and governed in practice — receive little attention in normative frameworks (Mittelsatdt, 2019; Leikas et al., 2019). A bottom-up approach thus does not merely trace the implementation of predefined principles; it can surface complementary and overlooked dimensions of governance that might otherwise go unrecognized.

The lack of user-centered perspectives in AI governance is also mirrored in the empirical literature. In their meta-review of 84 ethical AI documents, Jobin et al. (2019) observe that most ethical AI principles are developed without clear mechanisms for stakeholder engagement or implementation, potentially overlooking user perspectives and lived experiences. This finding echoes broader critiques of AI governance research, which has been characterized as conceptually rich but empirically thin (Wagner, 2018; Leikas et al., 2019). Despite a growing number of frameworks and guidelines, few studies systematically examine how governance is experienced or enacted "on the ground" by end users, employees, or managers (Stark & Hoffmann, 2019). As Morley et al. (2021) argue, there remains a disconnect between normative ideals and their translation into meaningful practice.

Together, these arguments point to the value of bottom-up approaches as a critical complement to existing top-down models. Such approaches help clarify how AI governance is shaped in practice — not only through policies and standards, but through the everyday concerns and adaptations of those who live and work with AI systems.

## 3. Methodology

To structure our analysis, Figure 1 provides a high-level overview of our methodological approach. We began by extracting topics from a large corpus of online reviews of AI products using BERTopic. These topics were then ranked according to their semantic similarity to two sets of governance-oriented reference texts: one based on established AI governance principles, and the other based on general, high-level governance terms. By identifying the sharpest semantic drop-off in similarity, we defined a cutoff point for selecting the most governance-relevant topics. Finally, we applied K-means clustering to these selected subsets to identify coherent thematic clusters of AI governance concerns as discussed by users.

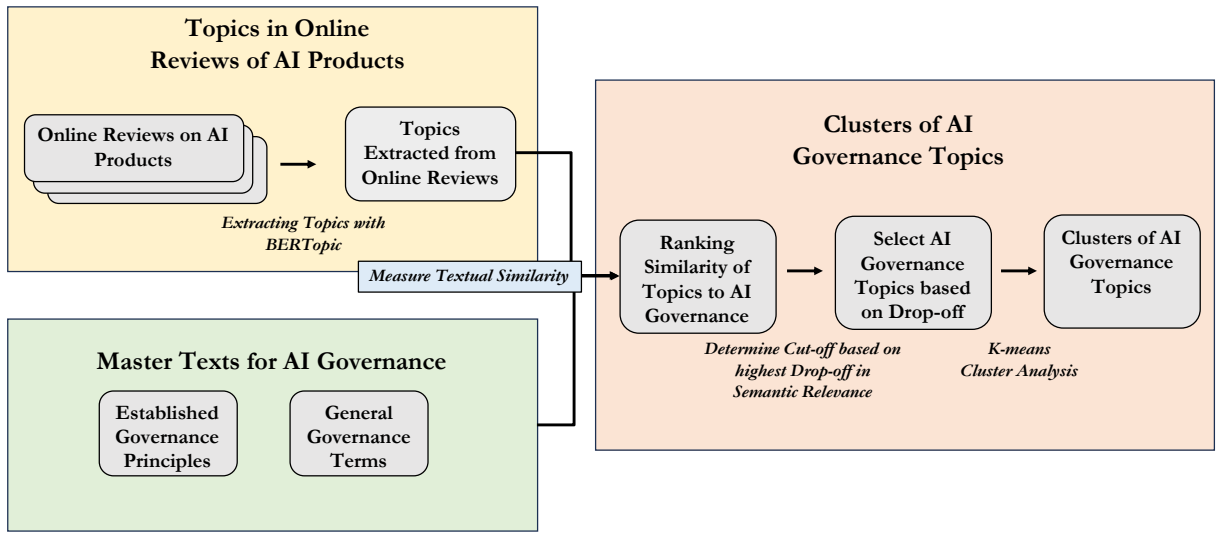

**Figure 1. Overview of Methodological Pipeline**

### 3.1. Data

Our analysis is based on user-generated reviews from G2.com, a leading platform for business-to-business (B2B) software evaluations. Unlike consumer-facing platforms such as Amazon or Trustpilot, G2 specializes in reviews of professional software tools, including a growing ecosystem of artificial intelligence (AI) applications (G2, 2024; Kevans, 2023). This focus on business-oriented use cases makes it a valuable source for examining how AI systems are experienced in practice, particularly in organizational or enterprise contexts. G2 data has recently been used to investigate Human-AI Interaction patterns in the workplace (Pasch & Ha, 2025), further highlighting its relevance for user-centered AI research.

We collected reviews from G2's "Artificial Intelligence" category, which encompasses a wide array of AI-driven products including chatbots, data science platforms, generative AI tools, and intelligent automation systems. To ensure topic robustness and data consistency, we included only products with at least 50 reviews. The resulting dataset spans 249 AI products, comprising a total of 108,998 individual user reviews.

### 3.2. Topic Modeling of User Reviews with BERTopic

To extract governance-related themes from user-generated reviews of AI products, we employed BERTopic (Grootendorst, 2022), a transformer-based topic modeling approach designed to capture nuanced patterns in unstructured text. BERTopic is particularly well-suited to analyze user discourse because it does not require a predefined number of topics and can flexibly adapt to the semantic structure of the data. Unlike more traditional methods such as Latent Dirichlet Allocation (LDA) (Blei et al., 2003) or K-Means clustering, which rely heavily on word co-occurrence or fixed cluster counts, BERTopic leverages Sentence Transformer embeddings to identify topics based on semantic similarity rather than lexical frequency.

This embedding-based approach allows BERTopic to group together conceptually similar terms even when they differ lexically — a critical advantage when working with user reviews, which often feature informal, varied language. Moreover, BERTopic supports soft clustering, assigning probabilistic topic memberships to each document. This enables the model to represent reviews that mention multiple themes (e.g., ease of use, pricing, and data privacy) without forcing a single-topic assignment. Such flexibility is essential in the context of AI governance, where concerns around autonomy, fairness, or security frequently coexist with broader product evaluations.

By capturing these overlapping themes, BERTopic allows us to disentangle governance-related discourse from more general product feedback and identify the distinct ways in which users express governance concerns in their own terms.

### 3.3. Anchoring AI Governance Concepts with Master Texts

Once BERTopic generated a full set of topics from the corpus of user reviews, we sought to identify those that were relevant to AI governance and related concepts such as ethical AI and trustworthy AI. Since no standardized taxonomy exists for AI governance from a user-centered perspective, we began by constructing a "governance master text" — a list of words and phrases related to AI governance. This list served as a reference point to evaluate how semantically similar each topic was to governance-related language, allowing us to identify and extract the topics that meaningfully engaged with AI governance themes.

Defining such a master text presents an inherent trade-off in how far existing literature and frameworks should be utilized to construct a reference for AI governance-related terms. Relying too heavily on established sources risks reintroducing a top-down perspective—anchoring the analysis in expert-driven categories and potentially overlooking how users themselves articulate governance-related concerns. Conversely, constructing the master text too loosely—without grounding it in recognized concepts—risks capturing vague or peripheral content and missing well-established concerns such as privacy, transparency, or accountability. The challenge, then, is to strike a balance: grounding the analysis in meaningful conceptual foundations while remaining open to the ways governance is expressed organically in user language.

To balance this trade-off, we adopted a dual master text approach: one drawing on commonly used concepts and principles from the AI governance literature, and another composed of deliberately high-level, broadly associated terms to avoid steering the analysis too strongly toward existing frameworks.

**Established Governance Principles:** One approach utilizes well-established concepts related to AI governance and ethics. For this, we drew on the seven core principles outlined in the European Commission's *Ethics Guidelines for Trustworthy AI* (2019)—a widely cited and influential framework in both policy and academic domains—and use the words for the seven core principles for our governance master text. These principles are: accountability, transparency, fairness, human agency and oversight, privacy and data governance, technical robustness and safety, and societal and environmental well-being. Similarly, Papagiannidis et al. (2023), in a scoping review of 48 studies on responsible AI, find that the dominant concerns and concepts discussed across the literature can be meaningfully organized under these same seven thematic pillars. This alignment makes the framework a strong candidate for structuring the principle-based master text. Moreover, while formally labeled as "ethics" guidelines, the EU's *Ethics Guidelines for Trustworthy AI* are used in practice to guide AI governance efforts—partly due to the absence of comparable, dedicated governance frameworks, but also because they articulate core principles that have become central to how AI governance is conceptualized, benchmarked, and implemented across sectors (Larsson, 2021).

This principle-based version served as a conceptually grounded reference point, anchoring the analysis in widely used terminology from both policy frameworks and academic literature on AI governance. While not exhaustive, it reflects the dominant language surrounding AI ethics and governance and allows us to assess the extent to which user reviews engage with or diverge from these established concerns.

**General Governance Terms**: To complement the more structured reference based on established governance principles, we constructed a second master text using high-level, abstract terms frequently associated with governance discourse in both scholarly and industry contexts. These included: "Governance", "Ethics", "Trustworthiness".

We intentionally avoided including more specific principles (e.g., privacy, fairness) to reduce the risk of steering the model toward predefined themes. Likewise, we excluded compound terms such as "AI governance" or "ethical AI." Since the text corpus already focuses exclusively on AI-related products, adding "AI" to the seed terms (e.g., "AI Governance" instead of just "Governance") would have introduced little additional signal, and might have inadvertently shifted the focus toward technical-oriented content.

This conceptually open framing allows us to observe whether and how users invoke governance-like ideas in their own terms, without being anchored to institutional or domain-specific language.

### 3.4. Measuring Textual Similarity to Identify Topics related to AI Governance

To identify governance-relevant topics within the full set of topics generated by BERTopic, we applied an embedding-based semantic similarity approach. Using the *all-mpnet-base-v2* model from the Sentence-Transformers library, we encoded each topic by averaging the embeddings of its top 10 keywords, providing a semantic representation of the topic's core content. In parallel, we embedded each of the two AI governance master texts as aggregated text vectors.

We then calculated cosine similarities between each topic embedding and each master text embedding. This produced a ranked list of topics based on their semantic proximity to governance-related language. Topics with higher similarity scores were considered more conceptually aligned with AI governance and were retained for further analysis.

This embedding-based filtering offers key advantages for our exploratory setup. Rather than relying on surface-level keyword matching, it captures deeper conceptual alignment—even when users express governance concerns in language that diverges from institutional or academic terminology. At the same time, it preserves interpretive flexibility, guiding the topic space without rigidly enforcing predefined categories—thus balancing theory-driven anchoring with sensitivity to user-centered discourse.

### 3.5. Selecting Topics Based on Semantic Relevance Drop-off

After calculating cosine similarity scores between each BERTopic-generated topic and the two governance master texts, we faced a critical methodological decision: how many of the top-ranked topics should be included in the governance-relevant subset?

To address this, we adopted a data-driven cut-off strategy based on the largest relative drop in similarity scores. The logic behind this method is to balance two competing priorities:

- Topical relevance: ensuring selected topics meaningfully align with governance discourse.
- Topical breadth: capturing a diverse set of governance concerns as expressed by users, including more implicit or less conventionally framed themes.

The largest relative drop in similarity identifies the point where the conceptual alignment between the topics and the governance language declines most sharply. Topics above this point exhibit relatively consistent semantic proximity to the master text, while those below it begin to diverge — either due to ambiguity, thematic drift, or generality.

To avoid overly narrow cut-offs driven by statistical outliers, such as the often very steep drop between the first and second most similar topics, we required a minimum inclusion of 10 topics. This constraint ensures that the resulting subset captures a baseline level of thematic diversity while still being guided by the semantic structure of the data.

There are also alternative approaches to selecting the cut-off for the topics to include, such as including a fixed number of top-ranked topics or applying a fixed similarity threshold. However, these approaches pose important limitations in this context.

First, selecting a fixed number of top topics (e.g., the top 20 or 50) is inherently arbitrary. It lacks grounding in the actual semantic distribution and risks either underfitting (by excluding relevant themes) or overfitting (by including weakly related or off-topic content). There is no theoretical consensus on the correct number of governance-related topics in user discourse, and thus no non-arbitrary way to fix this parameter in advance.

Second, applying a fixed similarity threshold (e.g., cosine similarity > 0.60) assumes that similarity scores are stable and directly interpretable across contexts—which they are not. Cosine similarity values are context-dependent, shaped by the embedding model, vocabulary, topic granularity, and dataset characteristics. As such, there is no principled way to define what constitutes "sufficient similarity" in a generalizable manner.

In contrast, our relative drop-off method is responsive to the actual shape of the similarity distribution. It provides a principled and context-sensitive cut-off that reflects where conceptual coherence meaningfully declines. This enables us to strike a balance between topical focus and coverage, aligned with the goals of interpretive and user-centered governance analysis.

Figures AF1 and AF2 in Appendix A illustrate this selection logic. AF1 shows the similarity distribution for the established principles master text, where the steepest relative drop occurs between topics 31 and 32 — suggesting that the top 31 topics form a semantically coherent and governance-aligned subset. AF2 displays the results for the general governance terms master text, with a comparable drop-off between topics 18 and 19. These inflection points were used as cut-offs to define the final topic sets. In both cases, we conducted robustness checks using a broader inclusion of the top 50 topics (see Section 4.5), which confirmed that additional topics introduced thematic redundancy or weakened conceptual coherence.

### 3.6. Clustering Governance-Relevant Topics

Once the governance-relevant topics had been identified based on their semantic similarity to the master texts, we applied a second layer of analysis to organize these topics into broader thematic categories. To do this, we used K-Means clustering, a widely used unsupervised learning algorithm that partitions data into a predefined number of clusters by minimizing intra-cluster variance. Again, each topic was represented as a dense semantic embedding, generated using the same *all-mpnet-base-v2* Sentence Transformer model applied earlier.

To determine the appropriate number of clusters (k), we followed standard practice and applied two widely used heuristics: the Elbow Method (based on within-cluster sum of squares, WCSS) and the Silhouette Score (Rousseeuw, 1987). Both methods are commonly employed in natural language processing and unsupervised topic modeling contexts (Onumanyi et al., 2022).

For both master text conditions, we observed a minor inflection point at k = 7 in the elbow curve and a local maximum in the silhouette score at the same point. This convergence supports the stability of a 7-cluster solution and guided our subsequent thematic interpretation of the governance-related topics. The graphs for both elbow and silhouette method for both master texts are in Appendix B.

# 4. Results

**Table 1: Topics and Cluster for Established Governance Principles Master Text**

| Cluster | Topics | |
|---|---|---|
| E1: Data & Analytics | E1.1. Database<br>E1.2. Analytics<br>E1.3. Dashboards & Reporting | E1.4. Metrics<br>E1.5. Data Visualization and Business Intelligence<br>E1.6. Data |
| E2: Project Management | E2.1. Project Management<br>E2.2. Project Tasks | E2.3. Productivity & Organization |
| E3: Deployment & Infrastructure | E3.1. Cloud Infrastructure | E3.2. Deployment, Orchestration & Infrastructure |
| E4: Human Factors & Organizational Structure | E4.1. Employees & HR<br>E4.2. Industry & Data Center<br>E4.3. Legal Issues<br>E4.4. Customer Relation Management (CRM) | E4.5. Contacts & Personnel<br>E4.6. Ecosystem<br>E4.7. Workspace & Agents<br>E4.8. Social Media |
| E5: Permission & Roles | E5.1. Permission, Role & Access Rights | E5.2. Admin/ Administration |
| E6: Privacy, Risk & Regulatory Compliance | E6.1. GDRP & Compliance<br>E6.2. Privacy & Operations<br>E6.3. Dependencies<br>E6.4. Restrictions | E6.5. Security<br>E6.6. Complaints<br>E6.7. Networks<br>E6.8. Accessibility |
| E7: Product Management & Strategy | E7.1. Product Management | E7.2. Roadmap & Strategy |

**Table 2: Topics and Cluster for General Governance Terms Master Text**

| Cluster | Topics | |
|---|---|---|
| G1: Recruitment | G1.1. Candidates, Screening, Hiring | |
| G2: Content & Information Quality | G2.1. News Generation & Verification<br>G2.2. Politics and Disinformation | G2.3. Content Quality |
| G3: Legal & Compliance | G3.1. Laws<br>G3.2. GDPR & Compliance | G3.3. Policy & Violations |
| G4: Privacy & Security | G4.1. Privacy & Security | |
| G5: Human & Workplace | G5.1. Privacy & Operations<br>G5.2. Human-AI Interaction | G5.3. Employees & HR |
| G6: Permission & Roles | G6.1. Permission, Rights and Roles | |
| G7: Communication & Engagement | G7.1. Decision Making<br>G7.2. Social Listening<br>G7.3. Social Media | G7.4. Internal Communication<br>G7.5. Sales & Marketing<br>G7.6. Support |

## 4.1. Bottom-Up Topics of AI Governance

Table 1 presents the resulting bottom-up AI Governance clusters based on *Established Governance Principles* (E), while Table 2 summarizes those derived from *General Governance Terms* (G). Each cluster represents a coherent thematic domain, with subtopics labeled based on dominant keywords and interpreted relevance.

For transparency and reproducibility, the full list of topics, including top keywords, is provided in Appendix C. These include all governance-relevant topics used for clustering, grouped by cluster membership and annotated with their corresponding top-ranked keywords.

The clusters generated using a seed list informed by established AI governance principles reveal a broad range of governance concerns expressed in user reviews. These concerns span both technical and non-technical domains and are closely tied to organizational governance structures and implementation practices.

On the technical side, we observe a coverage of themes that mirrors what is often referred to as the AI value chain or data pipeline (Heeks & Spiesberger, 2024; Munappy et al., 2020). This begins with foundational layers such as E3: Deployment & Infrastructure, which includes concerns around containerization, cloud architecture, and backend tooling — components often invisible to users but critical for system stability and control. From there, the analysis moves to E5: Permission & Roles, encompassing system-level governance mechanisms such as user permissions, administrative roles, and access rights. This cluster represents a technical yet user-facing intersection: it regulates who can interact with AI features and under what conditions. Further along the chain, we encounter E1: Data & Analytics, which aggregates topics such as dashboards, metrics, visualizations, and performance insights — all mechanisms by which users interpret and monitor AI systems. In tandem, E6: Privacy, Risk & Regulatory Compliance surfaces safety concerns themes rooted in technical system capabilities — including data protection, security protocols, networks, and system dependencies.

The non-technical clusters reflect substantial aspects of organizational governance and operationalization. For example, E2: Project Management includes topics related to task tracking, workflow coordination, and tool integration — areas where users interface with AI systems embedded in project-level structures. Similarly, E7: Product Management & Strategy includes concerns around product roadmaps, strategic alignment, and how AI capabilities evolve or are prioritized — highlighting strategic reflections not only on what AI does, but why it is being developed in particular directions.

Moreover, E4: Human Factors & Organizational Structure brings together a diverse set of topics, including internal communication, intranet platforms, human resources (HR), customer relations, and even organizational roles tied to AI adoption. While not reducible to a single function, this cluster captures the diffuse ways AI becomes embedded in the everyday structure of work. At the same time, it reflects how broader organizational environments—including industry norms, legal frameworks, and governance pressures—shape the conditions under which AI is introduced, managed, and experienced within institutions.

Together, the clusters derived from established governance principles show how principles of governance — such as privacy, control, and accountability — are made tangible in various tools and processes and are relevant across a wide range of business functions.

The clusters derived from general governance terms (G) cover many similar topics, though they emerged from a different entry point—anchored in just a few high-level keywords. Again, several topics relate to technical aspects of AI systems. G6: Permission & Roles includes concerns around administrative access, user rights, and control mechanisms. Moreover, G4: Privacy & Security centers on user data protection, sensitive data security risks, and trust in the underlying systems.

In addition to technical topics, the clusters include a diverse range of organizational and communicative themes. G1: Recruitment addresses governance questions in hiring and candidate screening processes. G5: Human & Workplace touches on the interaction between AI systems and employee-facing functions such as HR, operations, or support tools. G3: Legal & Compliance reflects concerns around institutional responsibilities, regulatory adherence and organizational accountability. G7: Communication & Engagement captures a broad range of user experiences related to customer service, internal communication, sales enablement, and marketing. Finally, G2: Content & Information Quality surfaces user concerns about the clarity, credibility, and potential misinformation embedded in AI-generated or AI-supported content.

Overall, these clusters reflect how general references to governance can ground a wide spectrum of concerns — from system control and data management to human interaction, communication, and the informational quality of AI-driven outputs.

## 4.2. Mapping Dimensions of Bottom-Up AI Governance

To better understand the relationships across the clusters derived from both master texts, we synthesized and mapped the resulting themes and topics. A key observation was that the identified clusters span both technical and non-technical domains—ranging from backend infrastructure and system-level controls to organizational processes and communication practices. Moreover, within both domains, the topics collectively reflect a full workflow or value chain. For technical clusters, this ranges from foundational infrastructure and system deployment to data processing, analytics, and user-facing outputs. For non-technical clusters, the workflow spans internal strategies, organizational processes, and project coordination through to external-facing concerns such as customer interaction, regulatory compliance, and public communication. To make sense of these patterns, we organized the clusters along two intersecting dimensions that capture key distinctions in how governance concerns are distributed across the dataset.

- Technical vs. Non-Technical: This axis distinguishes between concerns grounded in system-level infrastructure and architectural considerations (e.g., deployment, orchestration, permissions, analytics) and those focused on organizational, human, or managerial governance (e.g., HR, communication, project management).

- Internal vs. External Focus: This axis captures the degree to which a governance issue remains contained within the organizational or system boundary (e.g., internal workflows, employee tools, backend infrastructure) versus pointing outward toward customers, users, regulators, or society at large (e.g., privacy risks, communication channels, legal norms). While this internal–external distinction is relatively intuitive for non-technical clusters (e.g., internal HR versus external customer interaction), it is more nuanced on the technical side. Here, "external" refers to technical elements that are user-facing or subject to regulatory scrutiny—such as data reporting, security, or privacy—while "internal" includes deeply embedded infrastructure such as container orchestration or backend administrative controls. In this sense, the axis can also be interpreted, as a downstream–upstream gradient along the AI development and governance pipeline (Heeks & Spiesberger, 2024; Munappy et al., 2020).

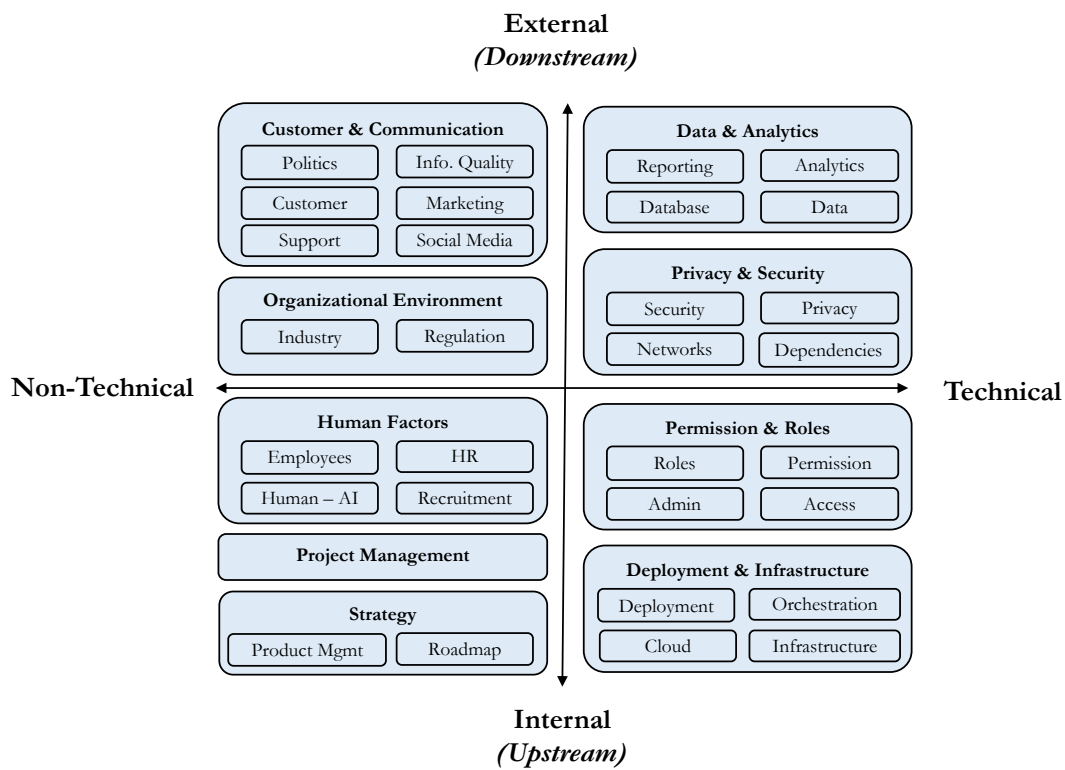

Figure 2: Summary of Bottom-Up AI Governance Themes

While these axes provide a useful heuristic for organizing governance concerns, the boundaries between categories are not always clear-cut. In many cases, organizational and technical dimensions are tightly intertwined. For example, data protection concerns under the GDPR reflect institutional and regulatory expectations, yet they are operationalized through system-level features such as encryption, user consent mechanisms, and access controls. This underscores that governance is often negotiated across multiple levels — from abstract principles to technical affordances — and that many themes exist on a continuum rather than fitting neatly into a single quadrant.

This matrix (shown in Figure 2) provides a conceptual overview of how users perceive and experience AI governance—not as a fixed institutional checklist, but as a distributed set of concerns shaped by infrastructure, human operations, external communication, and regulatory exposure.

The technical/internal quadrant captures foundational system-level operations that support the functioning of AI systems but remain largely invisible to end users. Clusters such as E3: Deployment & Infrastructure and E5 and G6: Permission & Roles address backend architecture, containerization, administrative control, and access rights—mechanisms through which technical governance is enforced within systems.

The technical/external quadrant reflects how AI systems manage privacy, security, and transparency in ways that are directed toward end-users and external stakeholders. Clusters like E6: Privacy, Risk & Regulatory Compliance and G4: Privacy & Security highlight user concerns around privacy, data protection, and system security — issues that directly affect customers and external trust. The quadrant also includes E1: Data & Analytics, which encompasses dashboards, performance metrics, and reporting tools that help make system behavior interpretable to both users and oversight bodies.

The non-technical/internal quadrant captures governance concerns situated within the internal operations and structures of an organization. Clusters such as E2: Project Management, G1: Recruitment, E4: Human Factors, G5: Human & Workplace, and E7: Product Governance & Strategy reflect how issues of oversight, accountability, coordination, and strategic alignment arise in the everyday management of AI systems. These themes highlight how governance is enacted through organizational roles, planning processes, and internal decision-making environments.

The non-technical/external quadrant encompasses governance concerns that extend beyond the boundaries of internal operations and reflect an organization's outward-facing responsibilities and contextual conditions. Clusters such as G7: Communication & Engagement, G2: Content & Information Quality, G3: Legal & Compliance, and E4: Organizational Environment point to how governance is negotiated through customer communication, public-facing messaging, and the credibility of AI-generated content. They also capture how broader industry conditions, legal frameworks, and societal expectations shape the governance landscape—highlighting the external forces that organizations must account for when deploying AI systems.

Together, this matrix constitutes a bottom-up governance map derived from user reviews—mapping the semantic structure of AI governance as articulated by practitioners, users, and stakeholders.

### 4.3. Comparing Established Governance Principles and General Governance Terms

The previous section presented a consolidated matrix of governance-relevant themes, derived from both general governance terms (G) and established governance principles (E). In this section, we examine how these two reference sets differ in the types of topics they surface and what this suggests about how different conceptions of governance influence the interpretation of user concerns.

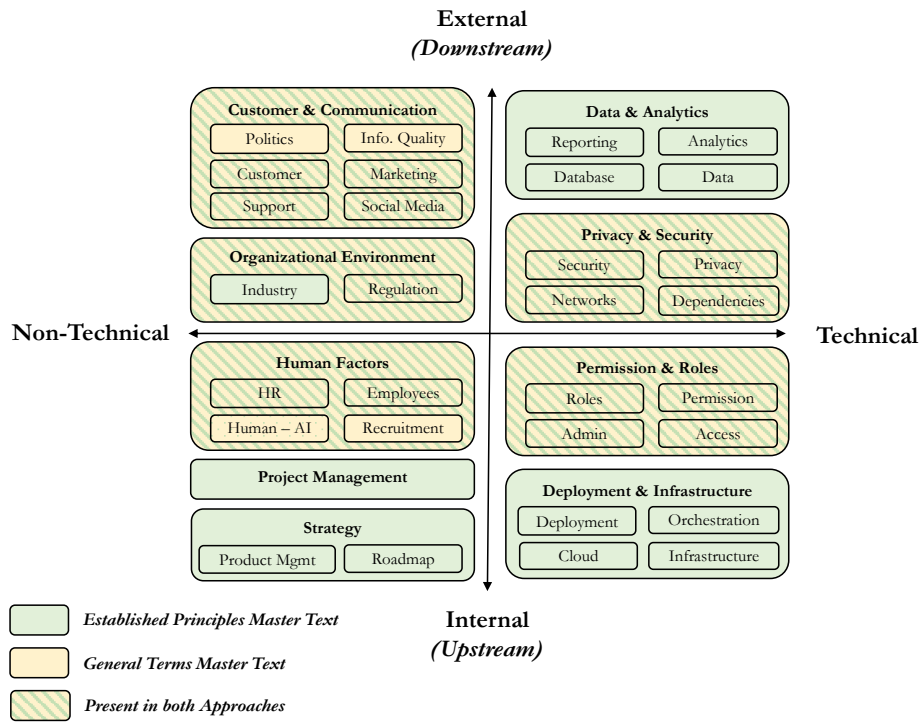

Figure 3. Bottom-Up AI Governance Themes by Master Text

Figure 3 highlights the distribution of AI governance themes by method. First, there is substantial overlap in the themes surfaced by the two master texts, suggesting that certain governance concerns are robust regardless of the methodological starting point. Shared themes include technical aspects such as privacy, security, permissions, and roles, as well as non-technical concerns like human factors related to HR and workplace dynamics. Additionally, both approaches surface communication-oriented themes—customer interaction, marketing, and social media—indicating a shared recognition of governance as it plays out in both operational systems and user-facing contexts.

At the same time, notable differences in emphasis emerge. The established principles approach (E) tends to highlight technical, operational, and managerial aspects of governance. For example, topics related to data analytics and deployment & infrastructure were unique to this method—consistent with the EU Ethics Guidelines' focus on system-level robustness and data governance (Larsson, 2021). Interestingly, this approach also surfaced clusters related to project management, strategic alignment, and industry context. While not explicitly defined in the principles themselves, these themes reflect how high-level values like accountability, transparency, and oversight are often enacted through organizational planning, coordination, and sector-specific adaptation in practice.

By contrast, the general governance terms approach (G) more frequently identifies clusters situated in the external, interactional domain, including Customer & Communication, Content & Information Quality, and Human Factors like Human-AI Interaction. These themes reflect how users experience AI governance through support systems, social media, messaging, and public interfaces. For example, clusters related to recruitment practices, public news, or internal employee communication emerged only when broader, high-level governance concepts were used to filter the topics.

Taken together, these findings highlight a complementary pattern. The established principles approach systematically recovers the backbone of what organizations and policy bodies tend to regulate — technical robustness, access control, and data governance. In contrast, general terms reveal the more relational, communicative, and experiential facets of governance as it is lived by users. The combination of both allows for a richer and more grounded representation of how governance manifests in practice — not just through regulation, but through organizational routines and everyday user interactions.

**4.4 Comparing Bottom-Up AI Governance Themes with EU AI Ethics Principles**

To contextualize our bottom-up governance clusters, we compare them with the widely cited *Ethics Guidelines for Trustworthy AI* proposed by the European Commission (2019). It is important to note that the match between bottom-up themes and top-down principles is not always one-to-one. While user discourse and institutional frameworks often describe governance in different vocabularies, they frequently touch on conceptually similar areas. For example, the principle of accountability may manifest in user concerns about admin rights, auditability, or access control—terms that do not appear in the guidelines but reflect similar underlying concerns.

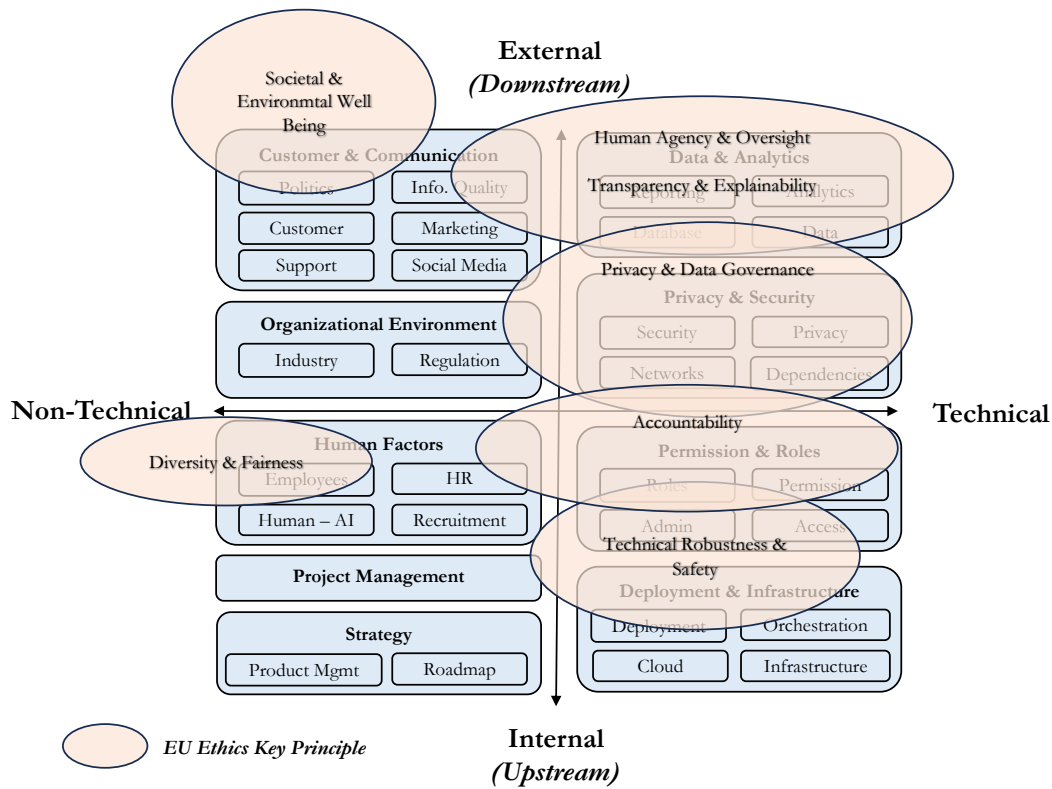

**Figure 4. Comparing Bottom-Up AI Governance Themes and EU Ethics Guidelines for AI**

Figure 4 visualizes this comparison by overlaying the EU's key principles onto the AI governance framework derived from user-generated reviews. This allows us to explore both areas of alignment and divergence between formal, top-down governance concepts and real-world, bottom-up concerns as voiced by users.

The visual mapping shows a strong degree of overlap in technical domains. EU principles such as Technical Robustness & Safety and Accountability correspond closely with user-identified clusters including Permission & Roles, Privacy & Regulation. These alignments suggest that privacy, security, access controls, and compliance mechanisms are recognized as governance concerns from both institutional and end-user perspectives.

We also observe conceptual alignment between principles in the EU ethics guidelines like Transparency & Explainability and Human Agency & Oversight, which overlap with clusters from our bottom-up analysis, such as Data & Analytics and Human Factors. This indicates that users are attuned not only to technical and procedural safeguards but also to concerns emphasized in governance principles, such as understanding system outputs, ensuring interpretability, and maintaining human oversight in AI-enabled environments.

However, notable differences emerge in the non-technical quadrants. User-derived themes like Customer & Communication, Internal Support, Product Strategy, and Project Management appear largely unaddressed by the EU's high-level principles. These clusters reflect more operational and interactional aspects of governance — such as how users experience AI in terms of support systems, roadmap visibility, recruitment practices, or marketing communications. These are critical touchpoints in AI deployment but are underrepresented in formal governance discourse. This disconnect reflects what Larsson (2021) has identified as a core limitation of ethical guidelines: an overemphasis on technical system features at the expense of sociotechnical and organizational dimensions. Moreover, while the EU guidelines broadly address technical concerns under high-level principles such as Technical Robustness & Safety, our findings suggest that user concerns often appear in more refined and operationalized terms—such as cloud, orchestration, and infrastructure. These granular dimensions of governance are not directly addressed in the EU's guidelines, pointing to a gap between abstract principles and the infrastructural realities of AI implementation.

Conversely, some of the EU's normative principles — such as Diversity & Fairness and Societal & Environmental Well-being — were not strongly reflected in the bottom-up themes. Their limited presence in user-generated reviews does not necessarily indicate disinterest, but rather suggests that users tend to express governance concerns in more immediate, task-related, or operational terms rather than abstract ethical language.

This pattern of topic overlap and divergence also extends beyond the EU's guidelines. Comparable analyses of other frameworks — such as the OECD AI Principles, which emphasize transparency, robustness, accountability, human-centered values, and sustainable development — and the meta-review by Jobin et al. (2019) — which highlights common values like fairness, responsibility, privacy, trust, and democratic participation — reveal similar trends. Core themes like privacy, transparency, and accountability are consistently represented across these high-level documents and show clear overlap with our user-derived clusters. However, these institutional frameworks also emphasize broad normative aspirations, including sustainability, democratic participation, and societal well-being, which rarely surface explicitly in user discourse. At the same time, they tend to underrepresent operational, organizational, and business-facing themes such as project workflows, customer interaction, and strategic product governance — areas that feature prominently in our bottom-up analysis.

Overall, the comparison underscores a key insight: while institutional frameworks tend to emphasize normative ideals and system-level safeguards, user discourse often reflects a situated, operational view of governance. Bridging this gap may require integrating high-level principles with practical mechanisms that align with how governance is experienced "on the ground."

### 4.5. Robustness Check: Fixed Topic Cut-off

To assess the robustness of our findings, we conducted a supplementary analysis using a fixed cut-off of the top 50 most semantically similar topics, rather than our primary drop-off-based threshold. We then applied the same K-Means clustering method, yielding 10 clusters each for the general governance terms (G) and established governance principles (E). This robustness check helps determine whether the main results are sensitive to the number of topics included or whether key governance themes remain stable across different thresholds. The full list of topics and cluster assignments from this robustness check is provided in Appendix D.

For both approaches, the results show a high degree of consistency with the earlier topic structures. Rather than introducing fundamentally new governance dimensions, the additional topics primarily enriched or extended existing clusters. For instance, in the general governance terms approach (G), the Recruitment & Hiring theme continued to form a distinct cluster, though now composed of two closely related topics (while previously consisting of a single topic), highlighting finer distinctions in candidate assessment and HR practices. Similarly, a broader and more granular set of topics emerged under "Communication & Engagement", now encompassing subtopics like support forums, sentiment analysis, and social media strategy.

On the established principles (E) side, the additional topics added more technical detail to the "Deployment & Infrastructure" and "Data & Analytics" clusters. For example, they include more finer-grained distinctions across APIs, orchestration platforms, and documentation tools. Similarly, the top 50 topics contains a variety of tools and practices around performance monitoring, databases, and dashboards—topics that were already conceptually present in the original clustering but are now more richly described.

Importantly, the broader structure of the clusters remained highly comparable. Most of the topics introduced at the 50-topic level aligned well with earlier clusters, either reinforcing or elaborating existing governance concerns. This lends strong support to the interpretive stability of the original framework and suggests that the cut-off point identified by relative drop-off already captured the majority of conceptually meaningful governance themes. The fixed-topics approach added granularity but did not substantially shift the thematic landscape.

## 5. Discussion

### 5.1 A Bottom-Up View of AI Governance

This study set out to explore how AI governance is discussed and experienced by users in applied, organizational settings — departing from the dominant top-down emphasis in the literature. Rather than starting from established principles or expert taxonomies, we examined user-generated reviews of AI products to identify which governance-related themes emerged organically from real-world use. By leveraging semantic similarity and unsupervised topic modeling, we derived a wide range of governance-relevant topics based on how users engage with AI systems in practice.

Our results reveal a diverse set of governance-related topics spanning both technical and non-technical domains. These themes cover a broad spectrum of organizational processes and decision structures — from strategic planning and internal coordination to customer interaction and compliance — as well as stages of the AI development pipeline, including deployment, data handling, and system control. To help illustrate how these concerns distribute across different areas of practice, we organize them into a 2x2 matrix (Figure 2), mapping their orientation along two key dimensions: technical vs. non-technical, and internal vs. external focus.

Crucially, this approach shows that governance is not just a matter of formal principles or abstract values, but a lived and operational phenomenon shaped by user-facing concerns and everyday interaction. The dimensions of governance surfaced in this study include practices, constraints, and decision points that may never appear in official guidelines, yet are central to how AI systems are evaluated, managed, and contested in practice.

By focusing on how governance appears in organic user discourse, this study contributes a descriptive and inductive perspective to the field. It highlights the value of seeing governance not only as a normative ideal, but as a set of topics, tensions, and strategies that emerge when AI technologies are used in practice. This bottom-up view complements existing frameworks, while challenging their assumptions about what governance should encompass. It also provides empirical evidence that can serve as a feedback mechanism for governance designers and policymakers — indicating which operational domains may require greater attention, support, or inclusion in future governance models.

### 5.2. Comparing Bottom-Up and Top-Down Perspectives

A key contribution of this study lies in contrasting user-derived governance themes with established top-down frameworks, such as the EU's *Ethics Guidelines for Trustworthy AI*. While these frameworks identify high-level principles — including transparency, privacy, and human oversight — our bottom-up approach reveals a broader set of concerns rooted in day-to-day operational and organizational practices.

Some areas show clear convergence. Topics related to privacy, security, legal compliance, and accountability appear prominently in both our results and institutional frameworks. This overlap reinforces the centrality of these concerns and suggests that top-down principles are not entirely detached from real-world user experiences.

However, our findings also highlight a set of governance-relevant themes that are largely absent in normative documents. These include project management, internal coordination, customer support, sales and marketing, and deployment infrastructure, all topics that emerged organically in user discourse but fall outside the traditional scope of AI governance and ethics guidelines. Their presence signals that governance, as experienced by users, often intersects with how AI tools are integrated into organizational routines and business functions, rather than just with normative ideals. This finding resonates with, for instance, Larsson's (2021) critique that many AI ethics guidelines privilege technical system features while neglecting the sociotechnical and organizational dimensions of governance. Our results empirically support this view: user concerns frequently reflect the embeddedness of AI in social, managerial, and communicative infrastructures — dimensions that rarely feature prominently in high-level policy frameworks.

This divergence underscores a key asymmetry in perspective. Top-down frameworks are typically constructed by policymakers and experts, often reflecting abstract ideals or regulatory aims. By contrast, bottom-up discourse foregrounds governance as situated practice — shaped by user interactions, organizational needs, and context-specific challenges.

At the same time, our analysis also highlights notable absences within user discourse. Several high-level governance principles — particularly those related to environmental sustainability, social and societal well-being, and fairness in a broader civic sense — were largely absent in the bottom-up topics. However, this absence should not be misinterpreted as a lack of concern. Rather, it reflects structural dynamics: users are often positioned to evaluate tools based on task performance, reliability, or organizational fit — not on systemic ethical outcomes. Ethical and environmental principles may be important at an institutional or regulatory level, but they do not always surface in day-to-day user interactions with AI systems. This highlights the need for multi-stakeholder approaches to AI governance that bring together macro-level ethical priorities and micro-level experiential concerns (Veale & Brass, 2019).

In this light, the two approaches serve different but complementary purposes. Top-down frameworks are essential for setting aspirational goals, ensuring regulatory accountability, and providing a shared ethical vocabulary. Bottom-up perspectives, by contrast, help uncover overlooked dimensions of governance and contextualize how principles are enacted — or contested — in applied settings. A comprehensive understanding of AI governance thus requires both: the normative guidance of high-level frameworks, and the empirical grounding of lived experience.

## 5.3. Implications for Governance Frameworks and Policy

The findings of this study carry important implications for the design and evolution of AI governance frameworks. As AI systems become increasingly embedded in everyday work environments, governance cannot be limited to high-level ethical ideals or technical safeguards alone. Instead, effective governance requires a closer alignment between abstract principles and the operational, organizational, and experiential realities of AI use.

First, the scope of governance frameworks should be broadened to reflect the practical dimensions of AI deployment. Themes such as project coordination, human–machine collaboration, internal communication, and customer support featured prominently in our bottom-up clusters. These concerns often fall outside traditional governance rubrics, yet they shape how AI systems are implemented, experienced, and adjusted in context. Recognizing such operational and business-oriented factors does not dilute ethical commitments; rather, it reflects how ethical considerations are embedded in and shaped by everyday organizational practice.

Second, our results highlight the need for a more inclusive understanding of governance stakeholders. Many top-down frameworks focus primarily on developers, policymakers, or auditors. However, the user-generated data analyzed in this study reveals that governance concerns also emerge from product managers, HR professionals, support teams, and end-users. These actors play an essential role in interpreting, adjusting, or even resisting how AI systems function — often in ways that top-down frameworks do not anticipate. Making these roles visible and involving such actors in governance deliberation and feedback loops, could improve the legitimacy and responsiveness of governance mechanisms. Approaches such as participatory standard-setting, user-informed benchmarking, and multi-stakeholder audits may help bridge the user-policy gap.

Taken together, these insights argue for a more integrated and iterative approach to governance, one that combines principled oversight with empirical attentiveness to everyday use. In doing so, AI governance frameworks can become not only more grounded, but also more responsive to the full range of actors, institutions, and interactions that shape the trajectory of AI in society.

## 5.4. Limitations and Future Research

While this study provides large-scale, bottom-up insight into how users discuss AI governance, several limitations should be acknowledged—both in terms of data and methodological scope—which also suggest directions for future research.

Using online reviews as the empirical basis allowed us to capture governance-related concerns as they are naturally expressed in everyday language. However, this text-based approach comes with inherent trade-offs. The data reflects the voices of a specific user group—primarily business and IT professionals active on platforms like G2—and does not include perspectives from other important stakeholder communities such as public sector actors, civil society organizations, or marginalized users. Moreover, while online reviews are useful for surfacing dimensions of governance that may be overlooked in formal frameworks, they offer limited insight into how such concerns are addressed in practice. Reviews rarely detail the challenges users face or the mechanisms they use to manage governance-related issues. Future research could address these gaps through complementary qualitative methods—such as interviews, ethnographies, or surveys—that provide richer contextual understanding.

It is also important to note that this study adopts a descriptive rather than normative perspective. Our goal was not to assess whether users' concerns align with ethical ideals, but to empirically map the dimensions of governance that emerge in bottom-up discourse. Accordingly, the absence of themes such as environmental sustainability, democratic accountability, or long-term societal impact should not be taken as an indication of their unimportance—only that they are not prominently discussed in the dataset. Future work may seek to explore whether and how such values can be incorporated into operational practice, and how institutional and user perspectives might be better reconciled.

Finally, the study is shaped by methodological and technical choices that warrant caution. The use of sentence-transformer embeddings, semantic similarity thresholds, and clustering algorithms such as BERTopic and K-Means inevitably entails abstraction and simplification. For example, the construction of governance master texts involves subjective design decisions that can influence which topics are highlighted. Alternative approaches—such as prompt-based analysis using large language models or participatory methods involving expert and user feedback—could refine and extend the robustness of the findings.

# Appendix

## Appendix A: Relative Dropoff in Semantic Similiarity

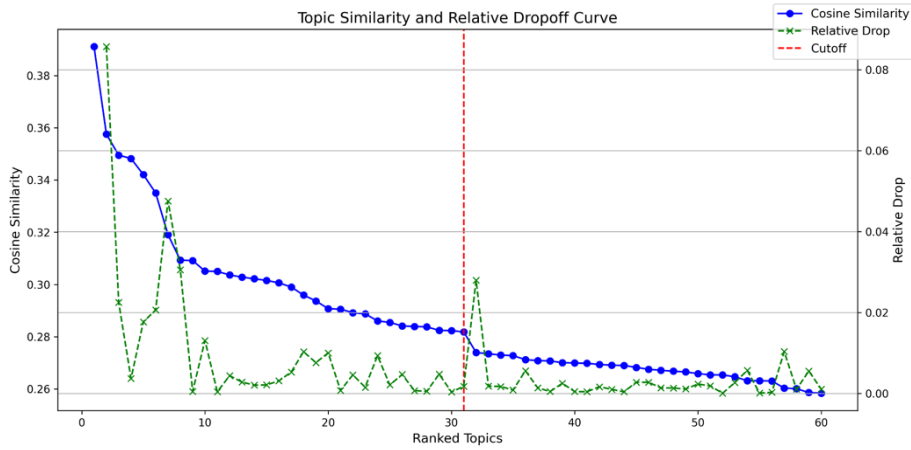

**Figure AF1: Relative Dropoff in Semantic Similiarity for Established Governance Principles Master Text**

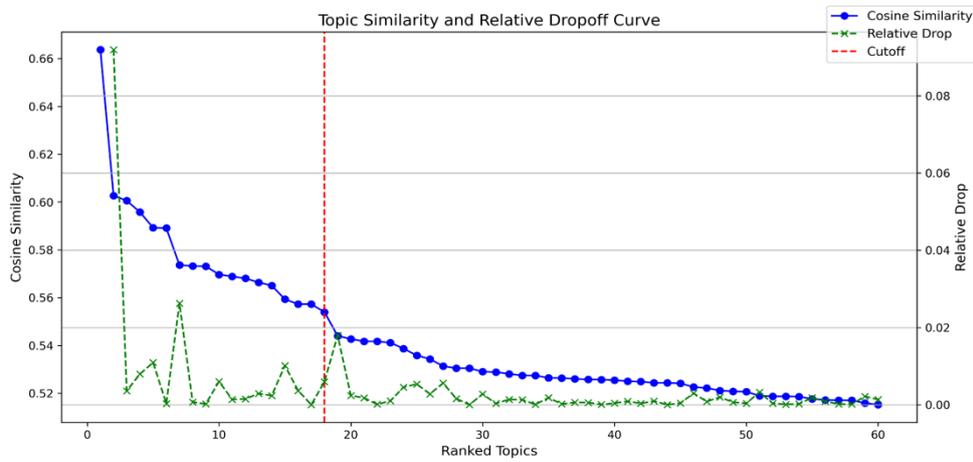

**Figure AF2: Relative Dropoff in Semantic Similiarity for General Governance Terms Master Text**

## Appendix B: Optimal Number of Clusters for AI Governance Topics

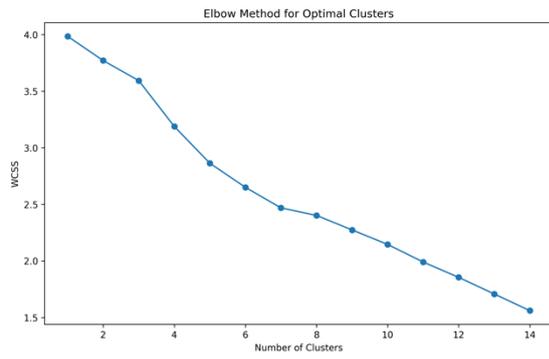

**Figure AF3. Elbow Method for Topics from Established Governance Principles Master Text**

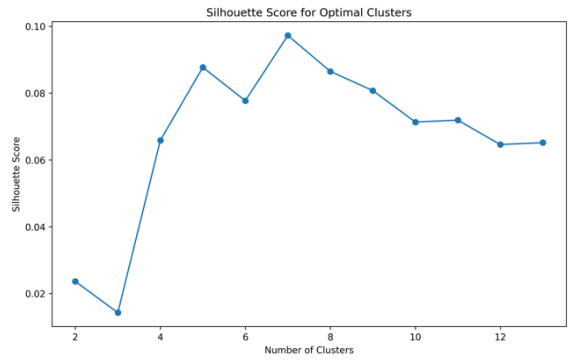

**Figure AF4. Silhouette Method for Topics from Established Governance Principles Master Text**

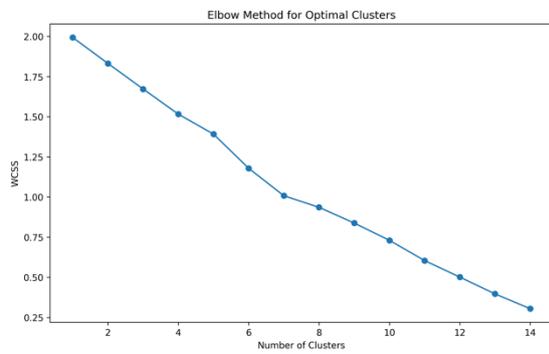

**Figure AF5. Elbow Method for Topics from General Governance Terms Master Text**

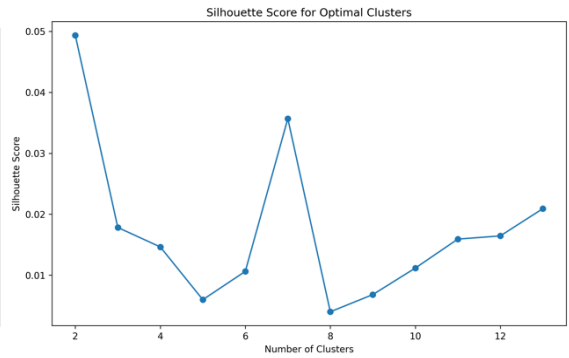

**Figure AF6. Silhouette Method for Topics from General Governance Terms Master Text**

## Appendix C: Topic List with Top Keywords per Topic

## Topic List for Established Governance Principles Master Text

**Cluster E1: Data Analytics**
- E1.1.:sql, database, databases, sources, mysql, db, connectors, relational, infer, queries
- E1.2.: analytics, stats, categorizations, advanced, metrics, dig, accessload, usageactivity, unresolve, analyticsit
- E1.3.: dashboards, dashboard, reports, reporting, customitation, puling, forethoughtagatha, viewsdashboards, resolutions, easytodigest
- E1.4.: metrics, dashboard, ingress, archival, trend, onetouch, salesforces, 24hour, adas, judge
- E1.5.: spotfire, tibco, visualizations, spotfires, visualization, bi, visuals, tableau, data, dashboards
- E1.6.: domo, domos, visualization, data, sources, visualizations, cards, dashboards, connectors, etl

**Cluster E2: Project Management**
- E2.1.: bitrix24, bitrix, management, 24, task, intranet, project, projects, tasks, activities
- E2.2.: hive, hives, project, projects, tasks, task, management, progress, gantt, views
- E2.3. notion, notions, notes, databases, productivity, organize, organizing, notetaking, workspace, pages

**Cluster E3: Deployment & Infrastructure**
- E3.1.: ibm, pak, cloud, governance, virtualization, manta, azure, containerization, multicloud, services
- E3.2.: truefoundry, kubernetes, docker, deployment, deployments, kubeflow, infrastructure, model, registry, orchestration

**Cluster E4: Human Factors & Organizational Structure**
- E4.1.: mebebot, employees, employee, carecom, hr, facilities, payroll, tier, questions, intelligent
- E4.2.: industries, industires, undersegmented, likebut, companiesindustries, datacenters, industry, coverage, avalaible, felxibility
- E4.3.: webhooks, webhook, illegally, technologywise, nonadmin, column, webservices, pabbly, hao, denied
- E4.4.: bitrix24, crm, bitrix, elitist, companys, management, pipelines, organizes, 24, cashstrapped
- E4.5.: contacts, lists, moniker, electionsgmailcom, couuld, calltoactions, ab, groups, workpersonal, categorize
- E4.6.: aggregate, industrybusiness, visualing, calcualting, complexcustom, ecosystemits, customerbrandcontact, easiliy, mobiledesktop, techies
- E4.7: agent, agents, workspace, supervisor, assist, verint, abusing, scripting, faqs, gladly
- E4.8: social, media, platforms, twitter, accounts, manage, place, facebook, networks, hootsuite

**Cluster E5: Permission & Roles**
- E5.1.: permissions, permission, roles, access, admin, rights, restriction, admins, control, administrator
- E5.2.: admin, administrative, administration, admins, macs, administrator, admining, activitiesappointment, buttonsnavigation, outinstead

**Cluster E6: Privacy, Risk & Regulatory Compliance**
- E6.1.: gdpr, europe, comply, compliance, privacy, burningwe, shook, regulartions, signupsignout, stroing
- E6.2.: workbot, workbots, knowledge, centralized, safe, operations, privacy, transformed, sensitive, efficiency
- E6.3.: dependence, relies, decisionmaking, raise, prescreening, privacy, perpetuate, quality, concerns, detrimental
- E6.4.: limitations, restrictions, limitation, constraints, env, regression, apaproved, urself, companyproject, outlineddefined
- E6.5.: privacy, protection, securiti, securitiais, sensitive, securitis, tumult, differential, masking, security
- E6.6.: complaints, complain, complaint, feautures, entrylevel, complaining, inconveniences, presets, aim, unavailable
- E6.7.: listening, social, particularities, uodates, mentions, managementlevel, networks, competion, comforting, listener
- E6.8.: convenience, ease, accessibility, practicality, versatility, interactivity, tastefor, bcozz, companiesalso, communicationradar

**Cluster E7: Product Management & Strategy**
- E7.1.: aha, roadmaps, ahas, jira, initiatives, goals, product, roadmap, managers, roadmapping
- E7.2.: roadmaps, roadmap, roadmapping, aha, strategy, devops, sprints, jira, softs, meaninful

# Topic List for General Governance Terms Master Text

**Cluster G1: Recruitment**
- G1.1.: glider, candidates, candidate, assessments, assessment, tests, recruitment, screening, hiring, fraud

**Cluster G2. Content & Information Quality**
- G2.1.: article, seconts, rapidity, authenticate, spinner, efficacy, articles, garbage, intelligently, resonates
- G2.2.: synthesia, synthesias, political, avatars, videos, disinformation, lifelike, actors, misinformation, independents
- G2.3.: decentquality, genrated, pumps, content, produces, friendliness, quality, seemless, shortform, originally

**Cluster G3. Legal & Compliance**
- G3.1.: contacts, hipaa, campains, definatly, affective, laws, decision, compliant, newbie, grouping
- G3.2.: gdpr, europe, comply, compliance, privacy, burningwe, shook, regulartions, signupsignout, stroing
- G3.3.: mistands, violate, butt, belong, intentions, specifics, how, know, policy, tell

**Cluster G4. Privacy& Security**
- G4.1. privacy, protection, securiti, securitiais, sensitive, securitis, tumult, differential, masking, security

**Cluster G5. Human & Workplace**
- G5.1.: workbot, workbots, knowledge, centralized, safe, operations, privacy, transformed, sensitive, efficiency
- G5.2.: ai, studios, artificial, intelligence, models, technology, forethought, simplified, peltarion, human
- G5.3.: mebebot, employees, employee, carecom, hr, facilities, payroll, tier, questions, intelligent

**Cluster G6. Permission & Roles:**
- G6.1.: permissions, permission, roles, access, admin, rights, restriction, admins, control, administrator

**Cluster G7. Communication & Engagement**
- G7.1.: dependence, relies, decisionmaking, raise, prescreening, privacy, perpetuate, quality, concerns, detrimental
- G7.2.: listening, social, particularities, uodates, mentions, managementlevel, networks, competion, comforting, listener
- G7.3.: social, media, platforms, twitter, accounts, manage, place, facebook, networks, hootsuite
- G7.4.: britix24, intranet, leaderships, internal, communication, communicate, intact, crossfunctional, talk, hence
- G7.5.: sales, marketing, signals, lead, prospects, leads, pages, enablement, growth, automation
- G7.6.: answers, interfaraince, eiiective, basehelp, questions, hackable, answering, orientate, owning, organizations

# Appendix D: Robustness Check with 50 Topics

## Topic List for Established Governance Principles Master Text with 50 Topics

**Cluster E1: Development**
- E1.1.: webhooks, webhook, illegally, technologywise, nonadmin, column, webservices, pabbly, hao, denied
- E1.2.: agent, agents, workspace, supervisor, assist, verint, abusing, scripting, faqs, gladly
- E1.3.: copilot, github, code, coding, developer, suggestions, developers, copilots, autocompletions, programming

**Cluster E2. Risk, Privacy & Organizational Friction**
- E2.1.: dependence, relies, decisionmaking, raise, prescreening, privacy, perpetuate, quality, concerns, detrimental
- E2.2.: limitations, restrictions, limitation, constraints, env, regression, apaproved, urself, companyproject, outlineddefined
- E2.3.: privacy, protection, securiti, securitiais, sensitive, securitis, tumult, differential, masking, security
- E2.4.: complaints, complain, complaint, feautures, entrylevel, complaining, inconveniences, presets, aim, unavailable
- E2.5.: listening, social, particularities, uodates, mentions, managementlevel, networks, competion, comforting, listener
- E2.6.: sales, marketing, signals, lead, prospects, leads, pages, enablement, growth, automation

**Cluster E3. Data Visualization & Business Intelligence**
- E3.1.: sql, database, databases, sources, mysql, db, connectors, relational, infer, queries
- E3.2.: spotfire, tibco, visualizations, spotfires, visualization, bi, visuals, tableau, data, dashboards
- E3.3.: domo, domos, visualization, data, sources, visualizations, cards, dashboards, connectors, etl

**Cluster E4. Privacy, Cloud, Infrastructure & Enterprise IT**
- E4.1.: gdpr, europe, comply, compliance, privacy, burningwe, shook, regulartions, signupsignout, stroing
- E4.2.: workbot, workbots, knowledge, centralized, safe, operations, privacy, transformed, sensitive, efficiency
- E4.3.: ibm, pak, cloud, governance, virtualization, manta, azure, containerization, multicloud, services
- E4.4.: mebebot, employees, employee, carecom, hr, facilities, payroll, tier, questions, intelligent
- E.4.5.: industries, industires, undersegmented, likebut, companiesindustries, datacenters, industry, coverage, avalaible, felxibility
- E4.6.: truefoundry, kubernetes, docker, deployment, deployments, kubeflow, infrastructure, model, registry, orchestration
- E4.7.: bitrix24, crm, bitrix, elitist, companys, management, pipelines, organizes, 24, cashstrapped
- E4.8.: coram, security, hardware, jasypt, buitifully, gelps, intrusions, encryptdecrypt, imporovement, multifactor
- E4.9.: docsie, documentation, docsies, technical, plugins, jwt, publish, knowledge, collaborating, clientside
- E4.10.: api, apis, documentation, apiscrapy, json, documented, docs, rest, washy, boxescoordinates
- E4.11.: cloud, gcp, google, scalability, releted, avilablity, manged, wrok, services, documention
- E4.12.: diversity, variety, incredibility, prerequisite, assesment, crowdsourcing, memorize, consist, inputs, knife

**Cluster E5: Project Management & Task Coordination**
- E5.1.: bitrix24, bitrix, management, 24, task, intranet, project, projects, tasks, activities
- E5.2.: hive, hives, project, projects, tasks, task, management, progress, gantt, views
- E5.3.: project, projects, task, tasks, management, view, kanban, gantt, status, assign
- E5.4.: hive, project, projects, duties, status, task, excelle, pageview, sharinng, adaptabilities

**Cluster E6: Social Media**
- E6.1.: social, media, platforms, twitter, accounts, manage, place, facebook, networks, hootsuite

**Cluster E7: Permission & Roles**
- E7.1.: permissions, permission, roles, access, admin, rights, restriction, admins, control, administrator
- E7.2.: admin, administrative, administration, admins, macs, administrator, admining, activitiesappointment, buttonsnavigation, outinstead

**Cluster E8: Communication & Interaction**
- E8.1.: contacts, lists, moniker, electionsgmailcom, couuld, calltoactions, ab, groups, workpersonal, categorize
- E8.2.: convenience, ease, accessibility, practicality, versatility, interactivity, tastefor, bcozz, companiesalso, communicationradar
- E8.3.: bing, chat, responsiveness, gpt4, chats, pertinent, enjoyable, microsoft, cap, restrict
- E8.4.: ease, implementation, frequency, maximal, cisco, okta, moderation, customer, integration, operational
- E8.5.: gemini, geminis, browsing, googles, surfing, privacy, security, tastes, browser, homepage
- E8.6.: britix24, intranet, leaderships, internal, communication, communicate, intact, crossfunctional, talk, hence

**Cluster E9: Product Strategy & Roadmapping**
- E9.1.: aha, roadmaps, ahas, jira, initiatives, goals, product, roadmap, managers, roadmapping
- E9.2.: roadmaps, roadmap, roadmapping, aha, strategy, devops, sprints, jira, softs, meaningful

**Cluster E10: Data & Analytics**
- E10.1.: analytics, stats, categorizations, advanced, metrics, dig, accessload, usageactivity, unresolve, analyticsit
- E10.2.: dashboards, dashboard, reports, reporting, customitation, puling, forethoughtagatha, viewsdashboards, resolutions, easytodigest
- E10.3. metrics, dashboard, ingress, archival, trend, onetouch, salesforces, 24hour, adas, judge
- E10.4. notion, notions, notes, databases, productivity, organize, organizing, notetaking, workspace, pages
- E10.5.: aggregate, industrybusiness, visualing, calcualting, complexcustom, ecosystemits, customerbrandcontact, easiliy, mobiledesktop, techies
- E10.6.: hana, sap, inmemory, erp, database, db, memory, cloud, s4, layer
- E10.7.: whylabs, observability, monitoring, monitors, whylogs, observatory, ml, ingestion, langkit, profiling
- E10.8.: alteryx, excel, data, workflows, tableau, workflow, community, repeatable, server, etl
- E10.9.: spss, modeler, ibm, statistical, prearranging, extensibilitypython, analyses, modeling, stata, syntax
- E10.10.: remotely, computers, remote, desktops, access, miles, security, traveling, control, device
- E10.11: dashboard, listener, chats, stats, convos, agent, count, analytics, customercreated, resolvedcreate

# Topic List for General Governance Terms Master Text with 50 Topics

### Cluster G1: Customer Communication & Internal Interaction
- G1.1.: britix24, intranet, leaderships, internal, communication, communicate, intact, crossfunctional, talk, hence
- G1.2.: customers, communicate, communication, interact, interaction, customer, reach, communicating, interactions, way

### Cluster G2: Communication & Engagement
- G2.1.: dependence, relies, decisionmaking, raise, prescreening, privacy, perpetuate, quality, concerns, detrimental
- G2.2.: listening, social, particularities, uodates, mentions, managementlevel, networks, competion, comforting, listener
- G2.3.: social, media, platforms, twitter, accounts, manage, place, facebook, networks, hootsuite
- G2.4.: sales, marketing, signals, lead, prospects, leads, pages, enablement, growth, automation
- G2.5.: answers, interfaraince, eiiective, basehelp, questions, hackable, answering, orientate, owning, organizations
- G2.6. overflow, stack, stackoverflow, teams, knowledge, public, questions, answers, private, qa
- G2.7.: sprinklr, sprinklrs, social, listening, sprinkler, media, moderation, care, channels, hence
- G2.8. funnel, strategies, marketing, punish, clientsbusinesses, rqeuirements, gathetring, ants, hightech, teachable
- G2.9. funcatinoality, simplicity, agnostic, studies, through, kinds, platform, run, functionalities, ease
- G2.10: conversocial, tray, mentions, conversocials, social, sentiment, publications, interactions, networks, comments
- G2.11.: soci, social, media, socis, posting, presence, posts, locations, franchise, platforms
- G2.12.: sentiment, analysis, sentiments, entity, textblob, detections, emotions, categorizing, detection, predicting
- G2.13.: briefs, seo, article, headings, bag, rankings, structural, labour, tom, guidance
- G2.14: terms, autopropagate, applicability, glossaries, finest, friendliness, communities, plethora, bases, man

### Cluster G3: Recruitment
- G3.1.: glider, candidates, candidate, assessments, assessment, tests, recruitment, screening, hiring, fraud
- G3.2.: candidates, sense, candidate, applicants, recruiters, screening, recruiter, talent, bullhorn, humanly

### Cluster G4: Regulation, Compliance & Coordination
- G4.1.: contacts, hipaa, campains, definatly, affective, laws, decision, compliant, newbie, grouping
- G4.2.: gdpr, europe, comply, compliance, privacy, burningwe, shook, regulartions, signupsignout, stroing
- G4.3.: set, independence, joe, challenging, difficult, concerning, initially, setting, intimidating, spreadsheet
- G4.4.: meeting, join, joining, meetings, 800number, asapp, quickjoin, convinience, invite, conduct
- G4.5.: contacts, lists, moniker, electionsgmailcom, couuld, calltoactions, ab, groups, workpersonal, categorize
- G4.6.: nonprofit, donors, nonprofits, donation, covid, hospitals, charity, gifts, grocery, contributions

### Cluster G5: Content & Information Quality
- G5.1.: article, seconts, rapidity, authenticate, spinner, efficacy, articles, garbage, intelligently, resonates
- G5.2.: synthesia, synthesias, political, avatars, videos, disinformation, lifelike, actors, misinformation, independents
- G5.3.: decentquality, genrated, pumps, content, produces, friendliness, quality, seemless, shortform, originally

- G5.4.: stuffysounding, rectfy, profitional, emails, writ, crispness, statements, emil, letter, mindless
- G5.5.: diversity, variety, incredibility, prerequisite, assesment, crowdsourcing, memorize, consist, inputs, knife
- G5.7.: intents, intent, entities, entity, utterances, verson, upserting, whichi, towards, tree
- G5.8.: polly, ssml, amazon, aws, voices, inflections, compliant, speech, natural, voice
- G5.9.: aspect, aspects, qlikautoml, hmm, couldnt, dislikes, encountered, wondershare, refer, downtime
- G5.10.: natural, sentiment, api, cloud, language, google, entity, analysis, automl, named
- G5.11.: convenience, ease, accessibility, practicality, versatility, interactivity, tastefor, bcozz, companiesalso, communicationradar
- G5.12.: rendering, easiness, quality, video, liked, gui, sound, accessibility, decent, before

**Cluster G6: Privacy & Security**
- G6.1.: privacy, protection, securiti, securitiais, sensitive, securitis, tumult, differential, masking, security

**Cluster G7: Complaints, Misconduct & Policy Violations**
- G7.1.: mistands, violate, butt, belong, intentions, specifics, how, know, policy, tell
- G7.2.: complain, complaints, fari, nothig, far, about, acts, road, professionalism, loving
- G7.3.: report, genrocket, fantastically, tanya, nothing, kinks, willingness, closely, handles, honestly
- G7.4.: neend, chatroom, cohosts, chat, scripted, reflecting, truthfully, anonymous, thrown, dislike
- G7.5.: nothing, meets, shot, cusomtization, managable, evil, sufficiently, honestly, recomend, collegues

**Cluster G8: Human-AI Interaction**
- G8.1. ai, studios, artificial, intelligence, models, technology, forethought, simplified, peltarion, human
- G8.2.: humantic, personality, disc, humanticai, prospects, cold, approach, buyer, traits, prospect
- G8.3.: hugging, face, pretrained, hf, models, nlp, sota, vision, huggingface, opensource
- G8.4.: avatars, avatar, clothing, clothes, outfits, cartoon, humanity, from, choice, choices

**Cluster G9: Permission & Roles**
- G9.1. permissions, permission, roles, access, admin, rights, restriction, admins, control, administrator

**Cluster G10: Workplace, HR & Finance**
- G10.1.: workbot, workbots, knowledge, centralized, safe, operations, privacy, transformed, sensitive, efficiency
- G10.2.: mebebot, employees, employee, carecom, hr, facilities, payroll, tier, questions, intelligent
- G10.3.: agent, agents, workspace, supervisor, assist, verint, abusing, scripting, faqs, gladly
- G10.4.: earnings, filings, financial, transcripts, financials, investor, reports, broker, equity, analyst